\documentclass[aps,prl,twocolumn,amsmath,amssymb]{revtex4}
\usepackage{graphicx}
\usepackage{bm}
\usepackage{color}
\usepackage{multirow}
\usepackage{times}

\begin{document}
\title{Quantum Melting of Valence Bond Crystal Insulators and Novel Supersolid Phase at Commensurate Density}
\author{Arnaud Ralko,${^1}$ 
Fabien Trousselet,${^{2,3}}$ and Didier Poilblanc$^{3}$ }
\affiliation{
${^1}$ Institut N\'eel UPR2940, CNRS 
and Universit\'e de Grenoble, F-38000 France \\
${^2}$ Max Planck Institut f\"ur Festk\"orperforschung, Heisenbergstrasse 1, D-70569 Stuttgart, Germany \\
${^3}$ Laboratoire de Physique Th\'eorique UMR5152, CNRS and Universit\'e de Toulouse, F-31062 France
} 
\date{\today}
\begin{abstract}
Bosonic and fermionic Hubbard models on the checkerboard lattice are studied
numerically for infinite on-site repulsion.  At particle density n=1/4 and
strong nearest-neighbor repulsion, insulating Valence Bond Crystals (VBC) of
resonating particle pairs are stabilized. Their melting into
superfluid/metallic phases under increasing hopping is investigated at $T=0K$.
More specifically, we identify a novel and unconventional {\it commensurate}
VBC supersolid region, precursor to the melting of the bosonic crystal.
Hardcore bosons (spins) are compared to fermions (electrons), as well as
positive to negative (frustrating) hoppings.
\end{abstract}
\maketitle

Understanding the conditions and mechanisms for charge conduction in strongly
correlated systems, especially frustrated ones, is a highly non-trivial
problem: aside from the strength of interactions, the lattice geometry, but
also statistics of charge carriers, can play a relevant role. A case of
interest is that of lattices with a pyrochlore-like structure: among those,
various unusual electronic phases are encountered, from (insulating) spin
liquids~\cite{herbertsmithite} to superconductivity, {\it e.g.} in spinels
involving $3d$ electrons as quarter-filled LiTi$_2$O$_4$~\cite{titanate}.
Also, other spinels like LiV$_2$O$_4$ have heavy-fermion behaviors~\cite{Fulde}
and are close to a Metal-Insulator (MI) transition~\cite{MI}, raising attention
to electronic correlations in a pyrochlore-based geometry. 

\begin{figure}[h] 
\includegraphics[width=0.45\textwidth,clip]{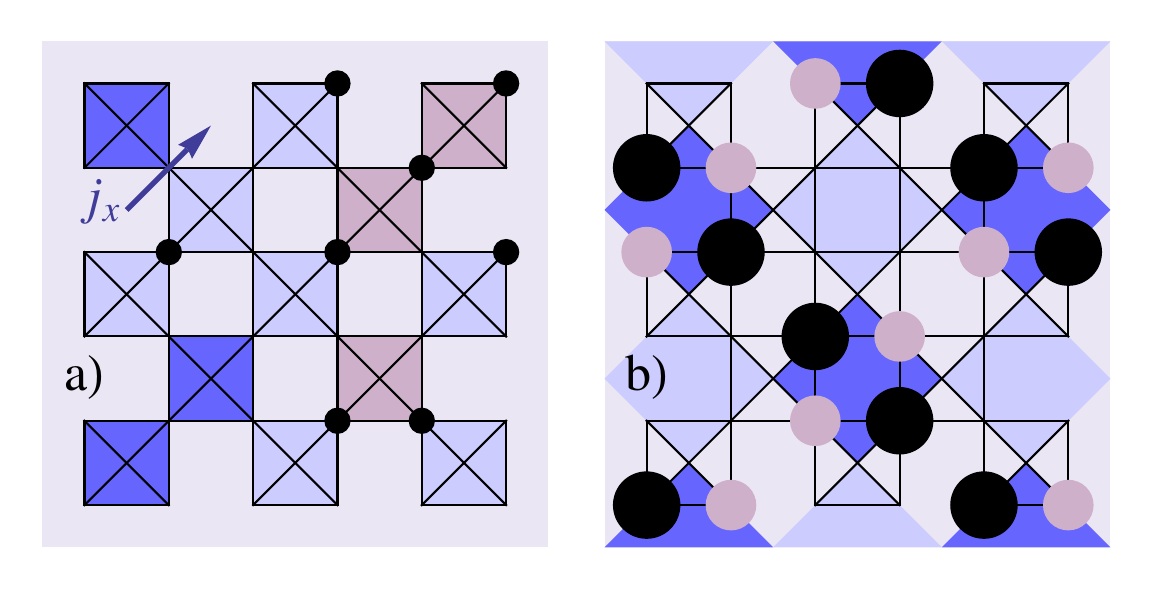}
\caption{(Color online).
a) Topological defects of fractional charge $+e/2$ (2 particles on a {\it red}
tetrahedron) and $-e/2$ (empty {\it blue} tetrahedron) and particle-hole
excitation with two pairs of NN defects. b) Artistic view of the { mixed-}VBC
supersolid phase at commensurate density $n$=$1/4$ showing inhomogeneous charge
density (dots), resonating plaquettes (blue) and superfluid density (light
blue).}
\label{Fig:defects}
\end{figure}

On the other hand, properties of some half-polarized antiferromagnetic
spinels~\cite{spinels} can be described similarly as those of anisotropic (XXZ)
magnets on a pyrochlore lattice. In the simplified case of a spin-1/2 XXZ
magnet, representing up- (down-)spin by the presence (absence) of a boson, a
mapping to hard-core bosons with NN repulsive interactions can be used; an
external magnetic field $h_z$, analog of a bosonic chemical potential, tunes
the bosonic density. At $h_z=0$ and only an Ising coupling on the bonds of the
checkerboard lattice of Fig.~\ref{Fig:defects} (the two-dimensional analog of
the pyrochlore lattice), the classical ground state (of density n=1/2) is
highly degenerate and follows the so-called {\it ice-rule} constraint: the
lowest Ising energy corresponds to precisely two bosons on every (flattened)
tetrahedron.  By tuning $h_z$ to reach a bosonic density of n=1/4, again the
classical ground-states obey an ice-rule constraint of precisely one particle
per tetrahedron (blue tetrahedra of Fig.~\ref{Fig:defects}). Second-order
processes in the exchange coupling lead to the dynamics of a loop
model~\cite{loop,heisenberg} (n=1/2) or of a quantum dimer model
(QDM)~\cite{rokhsar} (n=1/4). { Note that similar lattice gas picture can  also
be relevant to easy-axis spin-S systems~\cite{easy-axis} ($S\ge 3/2$).}

A specificity of these local constraints, aside from the finite ground state
degeneracy at the classical level, is that charge excitations may \textit{a
priori} be fractionalized~\cite{Fulde2}: adding a particle of charge $-e$ to a
system respecting initially the ice-rule creates 2 defects (tetrahedra on which
this rule is violated) of fractional charge $-e/2$ which can move to
neighboring tetrahedra (Fig.~\ref{Fig:defects}). An insulating phase with
fractional excitations could thus be considered, close enough to the MI
transition such that fractionalization is allowed by kinetic energy
gain~\cite{note-MI}.

Lastly, we note that supersolidity has been predicted under doping a
charge-ordered bosonic state~\cite{triangular}.  {\it Commensurate} supersolids
can be observed as an out-of-equilibrium state of cold
atoms~\cite{supersolid_of_equilibrium}, or arising from the melting of a
conventional charge-ordered insulator~\cite{supersolid2}. Recent developments
of {\it frustrated} optical lattices of cold
atoms~\cite{frustration_cold_atoms} open new directions.

To address the above issues, in this Letter we consider Hubbard hamiltonians on
the checkerboard lattice in the { infinite on-site repulsion (U)} limit that
reads: \begin{equation} \label{Eq:ham} H = -t \!\sum_{<i,j>,\sigma}\! P_G
(c_{i,\sigma}^+ c_{j,\sigma} + h.c.) P_G + V \sum_{<i,j>} n_i n_j,
\end{equation} where $c_{i,\sigma}^\dagger$ is a fermion { (electron)} creation
operator at site $i$, $\sigma$ { its} spin index  and $n_i $ is the onsite
density operator. The Gutzwiller projector $P_G$ enforces the $U$=$\infty$
limit (no doubly occupied site)~\cite{finite-U}. Note that the hopping term $t$
and the nearest neighbor (NN) Coulomb repulsion  $V$ have the same magnitude on
all bonds.  We also consider hardcore bosons (spins - a bosonic density $n<1/2$
corresponds to a spin system at polarization $1-2n$) replacing $c_{i,\sigma}$
by bosonic $b_i$ operators and dropping the sum on $\sigma$.

Although the lattice is frustrated, the $t>0$ bosonic model does not suffer
from a sign problem and can be addressed using Quantum Monte Carlo (QMC)
techniques.  In a closely related system, in which bosons hop on the square
lattice and interact via bonds on the checkerboard lattice, analytical
arguments and finite-T QMC simulations have shown that the system (extrapolated
to $T=0K$) undergoes, at density n=1/4, a {\it weakly first-order} phase
transition from a superfluid (SF) phase at small V/t to a large V/t MI
phase~\cite{SSE} (which breaks lattice translation symmetry), despite the
possibility of being an unusual non-Landau transition~\cite{Senthil}.  A
similar type of phase transition was also found on the Kagome
lattice~\cite{Kagome}.  Here, since the bosons can hop on {\it all bonds} of
the lattice, we performed Green Function QMC (GFQMC) simulations {\it directly}
at $T  = 0K$ and {\it in the canonical ensemble} (n=1/4) on clusters up to
$14\sqrt{2}\!\times\!  14\sqrt{2}$ (N=392 sites) to determine the emerging
quantum phases.  For frustrated $t<0$ and/or the case of fermions, studies are
restricted to smaller clusters, typically N=32 sites \textit{for a cluster
compatible with all lattice symmetries}) which can be handled using Lanczos
Exact Diagonalizations (ED)~\cite{poilblanc}.

\begin{figure}[h]
\includegraphics[width=0.45\textwidth,clip]{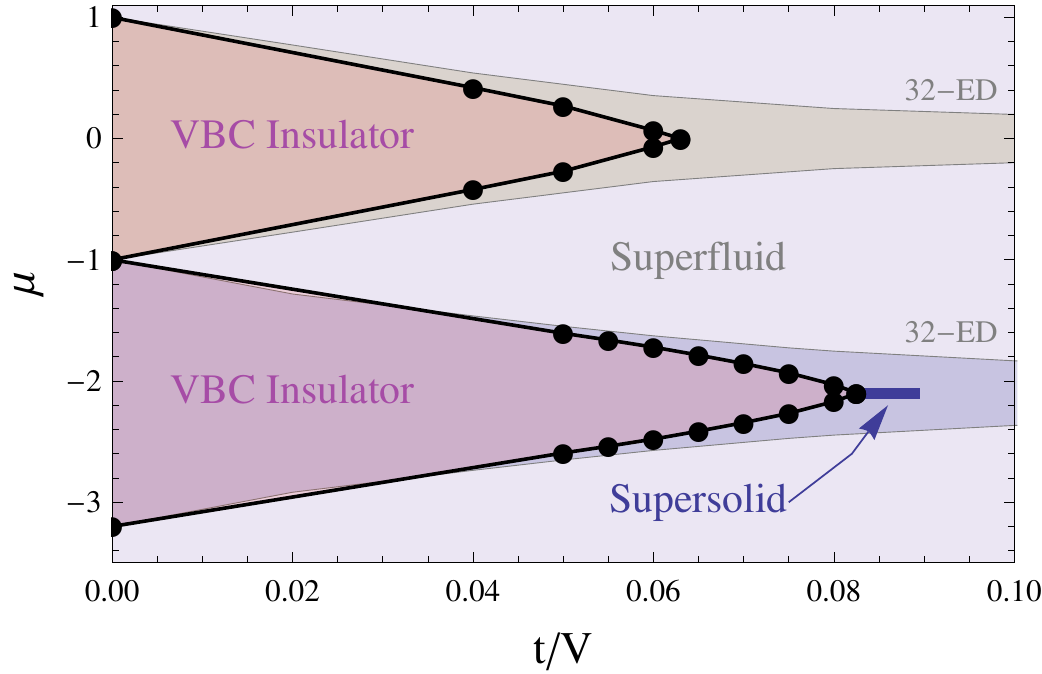}
\caption{(Color online). Phase diagram of the $t$$>$0 bosonic hard-core Hubbard
model as a function of the parameter $t/V$ and the chemical potential $\mu$.
Extrapolation to the thermodynamic limit (TDL) of $T$=$0K$ GFQMC results in the
canonical ensemble (dots) are compared to ED results. The two lobes correspond
to the n=1/4 (bottom) and n=1/2 (top) VBC insulators (see text). A region of
supersolid phase is found at commensurate $n$=$1/4$ filling (thick segment).}
\label{Fig:phase_diag}
\end{figure}

A typical phase diagram of eq.[\ref{Eq:ham}]  is shown in
Fig.~\ref{Fig:phase_diag} where the chemical potential $\mu$ has been
introduced via a simple Legendre transformation w.r.t. the density. This phase
diagram (for t$>$0 bosons, to be discussed later) raises a number of important
issues {\it i.e.} (i) the properties of the insulating VBC in the lobes, (ii)
the nature of the phase transition towards the itinerant phases, (iii) the role
of the quantum statistics (fermions vs bosons) and (iv) the effect of a
frustrating t$<$0 hopping. The paper is organized as follows.  First, we
investigate the evolution with increasing hopping of the gapped insulating
phase towards a compressible fluid. Next, we characterize the VBC insulator
from various relevant structure factors and discuss the presence of a novel
exotic supersolid phase with {\it coexisting VBC and SF orders}.  Third, optics
and dipolar excitations are discussed by calculating the optical properties in
the VBC insulators and in the itinerant metallic (fermions) or SF (bosons)
phases. Finally, we consider the physics of fractional defects beyond the
ice-rule constraint. Note that the results shown below refer to $n$=$1/4$, {\it
i.e.} to a quarter-filled bosonic (1/8-filled fermionic) system.

\begin{figure}[h]
\includegraphics[width=0.45\textwidth,clip]{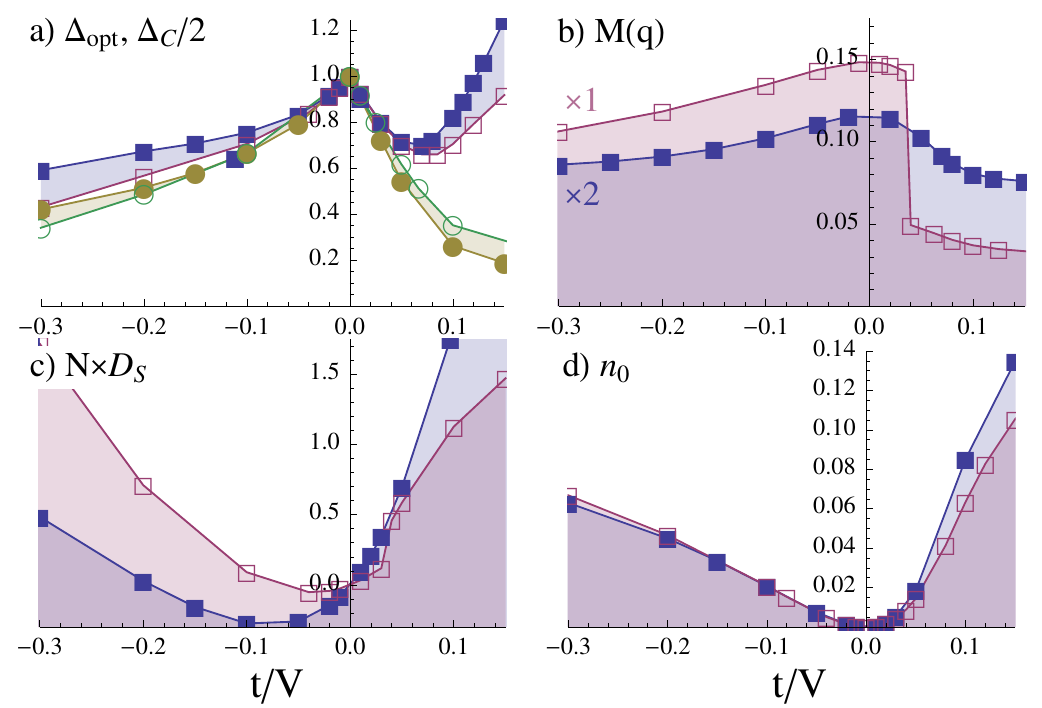}
\caption{(Color online). 
ED results obtained on a 32-site cluster for both fermions (empty symbols) and
bosons (filled symbols): a) charge (circles) and optical (squares) gaps, all in
units of $V$, b) order parameter $M_+ (K)$, c) Drude weight and d) average
number of $-e/2$ charge defects. For fermions, discontinuities due to a level
crossing at $t/V\simeq 0.035$ might be a finite size effect.} \label{Fig:ed}
\end{figure}

\begin{figure}[h]
\includegraphics[width=0.45\textwidth,clip]{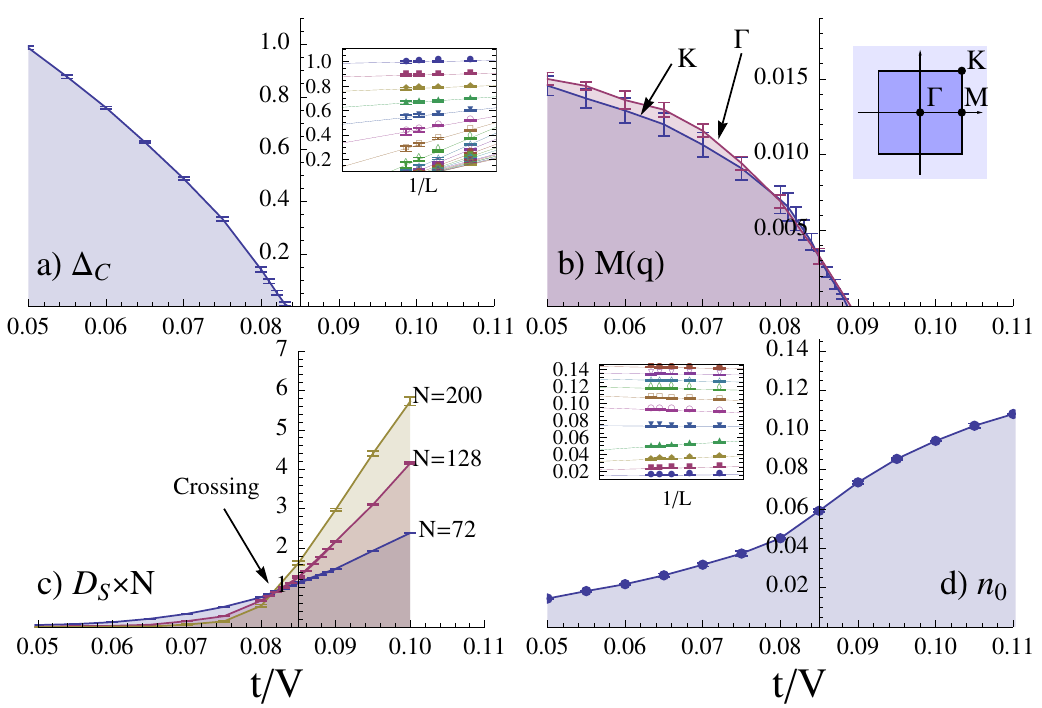}
\caption{(Color online).  GFQMC results for (hard-core) bosons with t$>$0. TDL
of the charge gap in units of $V$ (a), the order parameters $M_+(K)$ and
$M_-(\Gamma)$ with the Brillouin zone depicted in the inset (b), the stiffness
parameter rescaled by the number of sites $N$ (c) and the average number of
$-e/2$ charge defects (d). The same scale is used to plot these quantities
versus $t/V$ to emphasize the transition point close to $0.085$.  Size-scalings
are depicted in the insets.
\label{Fig:gfmc}
}
\end{figure}

{\it From the insulator to a compressible fluid} -- The insulating
(incompressible) character of a system manifests itself by a finite gap in the
quasiparticle charge excitation spectrum.  The gap closes in the metallic
(fermions) or SF (bosons) phase~\cite{note-MI}.  Physically, the process of
creating a particle-hole excitation as in Fig.~\ref{Fig:defects}(b) by moving a
particle from one edge of the sample to the other edge costs a finite energy
$\Delta_C$, the charge gap, simply calculated by $\Delta_C =  E_+ + E_- - 2E_0$
with the (total) ground state (GS) energies $E_0$ at filling $n$=$1/4$, and
$E_\pm$ with one extra (less) particle.  The ED results in Fig.~\ref{Fig:ed}(a)
show that, both for bosons or fermions, $\Delta_C$ decreases linearly with
$|t|$ from $2V$ at $t=0$ suggesting a transition at finite $t$ to an itinerant
GS.  Despite the small cluster size some trends can be noticed; (i) { the {\it
relative} robustness of the insulator w.r.t. its competing compressible phase
does not seem to differ significantly between bosons and fermions} and (ii) a
frustrating $t<0$ hopping is much less efficient to destroy the insulator (see
also Ref.~\onlinecite{poilblanc}). For unfrustrated bosons, GFQMC calculation
can be performed on much larger size clusters (up to 392 sites), at $T=0K$ and
for very small ratio of $t/V$. A careful size-scaling enables us to extract the
TDL of $\Delta_C$, displayed in Fig.~\ref{Fig:gfmc}(a).  The transition
characterized by the closure of $\Delta_C$ is located at $t/V\simeq 0.082$, not
so far from its estimation for the square hopping boson model~\cite{SSE}. Note
that the width $\Delta\mu$ of the n=1/4 and n=1/2 insulating lobes of
Fig.~\ref{Fig:phase_diag} is given by the extrapolated charge gap. 

{\it VBC insulators and supersolid (SS)} -- As discussed above, at large $V/t$,
mappings to generalized QDM describing the insulating phase are extremely
useful to establish the VBC nature of the insulator, in particular its {
feature of resonating particle pairs} at $n$=$1/4$~\cite{ralko,trousselet}.
For fermions, spin degrees of freedom play an essential role, since the
effective kinetic processes act only on {\it singlet} electron-pairs on the
void plaquettes~\cite{poilblanc,trousselet}. The insulators at $n$=$1/2$ and
$n$=$1/4$ have gapped triplet excitations and the same broken translation
symmetries~\cite{trousselet} as their bosonic VBC analogs of
Refs.~\onlinecite{loop} and \onlinecite{ralko}. However, when approaching the
phase transition by increasing $t$, the microscopic model (\ref{Eq:ham})
becomes necessary since the charge defects of Fig.~\ref{Fig:defects}(a) play a
crucial role.  We then compute the structure factor of the diagonal operator
$P_\pm = d_i d_j \pm d_k d_l$ defined in a previous work\cite{ralko}, where
$d_i d_j$ counts the pair (0 or 1) of particles facing each other across every
void plaquette (on the same plaquette at $\pi/2$ angle for $d_k d_l$). Note
that in the fermionic case, a projector on the 2-particle $S_z$=$0$ subspace is
implicit in the definition of $P_\pm$.  Following the same symmetry
considerations as in [\onlinecite{ralko}], the s-wave (d-wave) structure factor
at point $K$=$(\pi,\pi)$ ($\Gamma$=$(0,0)$) signals plaquette ordering
(rotational symmetry-breaking).  The related order parameters $M_+(K)$
(plaquette) and $M_-(\Gamma)$ (columnar)~\cite{note_nofq} are defined as
\begin{equation}
 M_\pm(q) = \sqrt{\frac{1}{N}\langle \psi_0 | P_\pm (-q) P_\pm (q)| \psi_0
 \rangle}.
\end{equation}
Data of Fig.~\ref{Fig:ed}(b) reveal striking differences between $t>0$ and
$t<0$ with a weaker suppression of the VBC plaquette order parameter with $|t|$
in the frustrated case, following the same trend as $\Delta_C$ vs $|t|$.  In
the bosonic $t>0$ model, a careful size-scaling analysis of QMC data again
enables to extract the TDL of the $T$=$0K$ order parameters as shown in
Fig.~\ref{Fig:gfmc}(b).  This evidences clearly that these two order parameters
of the mixed columnar-plaquette phase found in the large-V limit~\cite{ralko}
melt almost simultaneously at the phase transition around $t/V \simeq 0.089$.
Strikingly, the vanishing of the VBC order parameter appears at a larger value
than the insulator-SF transition.  This shows a persistence of the VBC order
within the SF phase, {\it i.e.} a small region of SS phase.  Note that (i) in
contrast to the bosonic triangular lattice~\cite{triangular}, this phase
appears here at a {\it commensurate} density,  (ii) its VBC character strongly
differs from usual charge ordering, and (iii) the presence of  kinetic
processes on the crossed bonds is essential, the VBC insulator-SF transition
being weakly first-order otherwise~\cite{SSE} with {\it no} intermediate phase.
{ However the {\it exact} equality between the bond strengths within the
crossed plaquettes is not crucial and the exotic SS phase should exist in an
extended parameter region.} We speculate that, for fermions, a similar (narrow)
exotic VBC metallic phase might also appear.

\begin{figure}[h] \includegraphics[width=0.45\textwidth,clip]{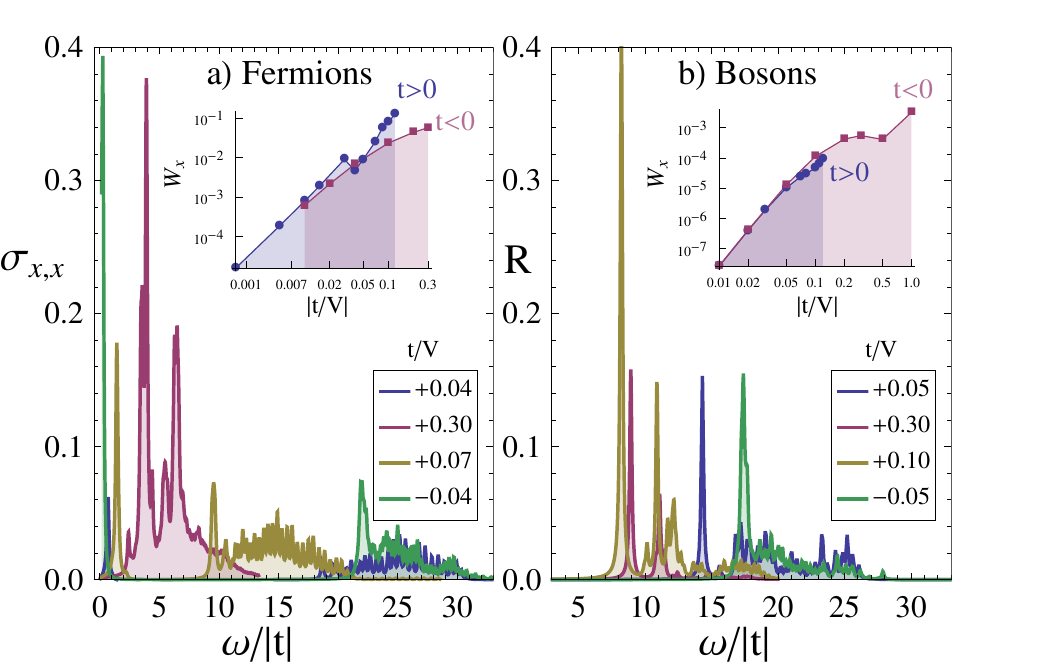}
\caption{\label{Fig:cond} (Color online). $\sigma(\omega)$ for a) the fermions
and b) the bosons at different values of $t/V$ (ED on a 32-site cluster).
Insets: relative weight of the low-energy dipolar matrix elements (only visible
for fermions on the $\sigma(\omega)$ plot), plotted vs $t/V$ in log-log scale.
} \end{figure}

{\it Optics} -- We now analyze the optical conductivity: first, it provides
important information on transport, from its $\omega\rightarrow 0$ limit, the
Drude weight (its finiteness characterizing a metal or SF).  Note that for
bosons at $T$=$0K$, this Drude weight corresponds to the stiffness of
superfluidity~\cite{capello}. Secondly, {\it neutral dipolar} excitations
should appear in the finite frequency absorption spectrum.  Observe that, for
fermions, these excitations are also spin singlets.  First, we calculate by
Lanczos ED the finite frequency conductivity, \begin{equation}
\sigma_{x,x}(\omega)=\frac{1}{\omega}{\rm Re}\big<\psi_0|
j_x\frac{1}{\omega+i0^+-H+E_0}     j_x|\psi_0\big>\, , \label{Eq:cond}
\end{equation} where $j_x$ is the current operator along one of the diagonal
directions as shown on Fig.~\ref{Fig:defects}.  The data on Fig.~\ref{Fig:cond}
clearly show, at small $|t|/V$, an optical gap $\Delta_{\textrm{opt}}\sim V$
corresponding to an absorption energy threshold.  As shown in
Fig.~\ref{Fig:ed}(a), this gap follows closely - for small enough $|t|/V$ - the
linear behavior with $|t|$ of $\Delta_C$. We believe the up-turn at $t/V\sim
0.08$ indicates the transition to the metallic/superconducting phase, where the
energy scale of excitations cross-over from $V$ to $t$.

Interestingly, low-energy peaks are observed in $\sigma(\omega)$ {\it below}
$\Delta_\textrm{opt}$ at energies $\sim$$t^2/V$ set by the effective QDM
hamiltonian.  While such $q$=$(0,0)$ low-energy dipolar excitations correspond
to usual excitations of a metal/SF, it is not clear yet whether they would also
survive in the TDL in the insulator, as occuring from local charge
fluctuations.

The Drude weight $D_S$, extracted from the above ED results via the optical sum
rule and plotted in Fig.~\ref{Fig:ed}(c), increases rapidly with $|t|/V$ but
does not allow to locate the insulator-metal (or SF) transition for such small
systems.  Fortunately, in the case of bosons and t$>$0, it is possible to
compute the Drude weight with the GFQMC method using the winding numbers of the
flow of the particles in imaginary time $\tau$~\cite{capello}.  At a continuous
solid to SF transition, the stiffness scales as $D_S = L^{-z} f(
L^{1/\nu}\left( t/V - (t/V)_c \right) , \tau L^{z} )$ with the dynamical $z$
and the correlation length $\nu$ exponents\cite{Kagome}. Searching for a
universal behavior of $f$ w.r.t. $L$, we obtain a very accurate estimation of
the transition at $t/V \simeq 0.082$ (crossing in Fig.~\ref{Fig:gfmc}(c)), with
$z=2$ and $\nu=1/2$. In agreement with the vanishing of $\Delta_C$, this is an
extra evidence in favor of a window of width $\Delta (t/V)\simeq 7\cdot10^{-3}$
of a SS phase exhibiting a finite SF stiffness $D_S$ and $\Delta_C =0 $ but
with { two} finite VBC order parameters.

{\it The physics of fractional defects} -- Within the effective QDM, the
amplitude of dimer flips ({\it i.e.} simultaneous 2-particle move on a void
plaquette~\cite{loop}) goes like $t^2/V$.  Hence, the concentration $n_0$ of
$\pm e/2$ defects ({\it cf} Fig.~\ref{Fig:defects}(a)) shows in
Fig.~\ref{Fig:ed}, for very small $t$, a $(t/V)^2$ behavior, symmetrically for
$t>0$ and $t<0$.  However, as soon as $|t|/V > 0.03$, the observed behaviors
are beyond the physics of the effective QDM.  Nevertheless, the phase
transition occurs at a small $t/V$ value so that the ice-rule constraint should
still operate in the metallic/SF phase which contains only $\sim$6-10$\%$ of
fractional defects.  Since the deconfinement~\cite{SSE} of these defects occurs
at the vanishing of the VBC order parameter, they should still be confined in
the SS phase.

We summarize here the main findings of this work.  First, a strong asymmetry is
found between t$>0$ and t$<$0, a frustrated hopping (t$<$0) being less
efficient to melt the VBC insulator. This originates from a relatively smaller
increase with $|t|$ of the average kinetic energy and concentration $n_0$ of
fractional charged defects, as shown in Fig.~\ref{Fig:ed}(d). Secondly, for the
non frustrated $t$$>$0 bosonic model, evidence are provided for a {\it double}
quantum phase transition between the VBC insulator at small $t$ and the SF at
large $t$, involving an intermediate {\it commensurate}~\cite{theorem}
supersolid phase breaking {\it both} translation {\it and} rotational
symmetries. Its VBC character (anisotropic resonating plaquettes) clearly
distinguishes it from its charge ordered counterpart~\cite{supersolid2}. We
speculate that such a feature should be present in the frustrated t$<$0 case
and/or for fermions (with an intermediate VBC metal).  Lastly, optical
properties are investigated, showing an optical gap following the behavior of
the charge gap.

{\it Acknowledgements} -- We thank K. Damle, P. Fulde, N. Laflorencie and S.
Wessel for interesting discussions.  F.T. is grateful to IDRIS (Orsay, France)
for computer time.

 
\end{document}